**Dispersion and losses of plasmons along simple metasurfaces: the analysis of dispersion equations**

Michael V. Davidovich

*Department of Radio Engineering and Electromagnetics, Institute of Physics, Saratov National Research State University, 410012, Saratov, Russia*

Corresponding e-mail: davidovichmv@info.sgu.ru

Michael V. Davidovich is Professor in Saratov National Research State University.

# Dispersion and losses of plasmons along simple metasurfaces: the analysis of dispersion equations

The flat metasurfaces described by tensor surface conductivity, the transverse size of which is small compared to the wavelength, are considered. In this case, we introduce two-dimensional surface conductivity for them, as well as for infinitely thin conductive sheets of graphene type. The method of Green's tensor functions of electrodynamics connecting fields and current densities, as well as the mode matching technique, are used. Conductive films on substrates are considered, including layered substrates of finite thickness with periodic layers and infinite substrates, as well as gradient substrates with a dependence of the dielectric constant on the thickness. Two-dimensional periodic structures of conductive films and dielectric films doped with nanoparticles are also analyzed. The possibility of applying the method to diffraction of surface plasmons on metasurface inhomogeneities is analyzed.

## 1. Introduction

Recently, much attention has been paid to metasurfaces and plasmon polaritons (PP) along them [1–12], since structures with surface PP have very wide areas for applications [13–35]. In a number of works, metasurfaces are considered, during the passage of which light changes the phase, while the Snellius law is changed [13]. This is due to the resonant interaction with meta-atoms on the surfaces, which are considered as surfaces with a finite thickness. We do not consider such effects, limiting ourselves to thin compared to the wavelength metasurfaces, the dispersion equations (DE) for which can be obtained by stitching fields. We limit our consideration to the study of surface plasmon polaritons along the simplest meta-surfaces that can be considered as a 2D electron gas (2DEG), such as graphene. We consider the metal film to be thin and as 2DEG if its thickness is much smaller, then wavelength and the depth of field penetration. These are usually thicknesses of the order of nm, but not more than a few dozen nm. We consider such simple metasurfaces described by complex surface tensor conductivities $\hat{\sigma}$, with an emphasis on taking into account dissipation. Dissipation, even weak, is very important and significantly affects the properties of surface PP. We use a generalized approach, considering surface PP with both the surface character of the field and its leakage (anti-surface) character. In this sense, the term surface PP means that we are interested in propagation along the surface. The simplest such surface is graphene. If its conductivity $\sigma(\omega)$ is considered as scalar without spatial dispersion, then, as is known, there are two PP with complex decelerations $n(\omega) = \sqrt{1 - 4/(\eta_0 \sigma(\omega))^2}$ and $n(\omega) = \sqrt{1 - (\eta_0 \sigma(\omega))^2/4}$, $\eta_0 = \sqrt{\mu_0/\varepsilon_0}$ for E- and H-plasmons [36]. The presence of a dielectric constant (DP) substrate with a thickness of $d$ leads to decelerations $n(\omega) = \sqrt{1 - 1/(\eta_0 Y_{in}^e(\omega))^2}$ for the E-plasmon along the $x$ axis, and $n(\omega) = \sqrt{1 - (\eta_0 Y_{in}^h(\omega))^2}$ for the H-plasmon. Here

$$y_0^{e,h} = -Y_{in}^{e,h} = y^{e,h} \frac{y_0^{e,h} + i y^{e,h} \tan(k_z d)}{y^{e,h} + i y_0^{e,h} \tan(k_z d)} + \sigma \eta_0, \qquad (1)$$

$y^e = k_0 \varepsilon / k_z$, $y^h = k_z / k_0$, $k_z = \sqrt{k_0^2 \varepsilon - k_x^2}$, and values with index 0 are obtained at $\varepsilon = 1$. To obtain the dispersion equations, it is more convenient to express $y_0^{e,h}$ as

$$y_0^{e,h} = \frac{\sigma \eta_0}{2} \pm \sqrt{\frac{(\sigma \eta_0)^2}{4} + \frac{y^{e,h} \sigma \eta_0}{i \tan(k_z d)} + (y^{e,h})^2}. \qquad (2)$$

Then we get a different form $Y_{in}^{e,h}$ for these equations. Although (1) and (2) are equivalent, the right-hand sides of (2) do not use the conductivities $y_0^{e,h}$ through which the decelerations are determined. From (2) it is possible to determine the conductivities $y^{e,h}$:

$$y^{e,h} = -\frac{\sigma \eta_0}{2i \tan(k_z d)} \pm \sqrt{(y_0^{e,h})^2 - \sigma \eta_0 y_0^{e,h} - \left(\frac{\sigma \eta_0}{2 \tan(k_z d)}\right)^2},$$

from where we get the form DE $n^h = \sqrt{\varepsilon - (y^h)^2}$, $n^e = \sqrt{\varepsilon - (\varepsilon/y^{e,h})^2}$. Note that all the equations are implicit, i.e. the right-hand sides depend on frequency and decelerations $n^{e,h}(\omega)$. Next, we will consider the graphene model with SD [37–42], although PPs in graphene are usually considered without taking into account SD [43–46], since it is weak up to the THz range.

The purpose of the work is to consider the properties of surface waves or PP along the simplest metasurfaces, methods for calculating PP dispersion and losses with taken into account the spatial dispersion (SD). We mean by such a metasurface a flat surface on which it is possible to introduce the input surface tensor impedance. By such simple metasurfaces we also mean the atomically thin sheet such as a graphene monolayer, graphene bilayer, borophene, or a thin layer with a thickness $h$ much less than the wavelength ($h \ll \lambda$), configured in a certain way from a conductive and/or magnetodielectric thin film and located on a substrate. Such substrate is described by the dielectric permittivity (DP) $\tilde{\varepsilon} > 1$, the dispersion and losses in which we do not take into account.

In this paper, we develop such approach based on the electrodynamic tensor Green's functions (GFs) for metasurfaces described by arbitrary tensor surface conductivity. Such conductivity can be, for example, formed by a thin metal film perforated in the form of a two-dimensional lattice. A thin conducting layer can be considered as two-dimensional electron gas (2-DEG), and in the case of a 2D-periodical structure it is described by tensor 2D (surface) conductivity $\hat{\sigma}$. The methods for obtaining such conductivity are known [39–46] and will be discussed further. 2-DEG, for example, occurs in the case of a metal and semiconductor film, or in the case of a graphene sheet. A periodically perforated metal film and graphene with etched holes can correspond to tensor conductivity, as well as a graphene sheet periodically doped with certain atoms and periodically graphene nanoribbons. The graphene sheet itself has weak anisotropy and tensor conductivity at high frequencies, where SD is significant [42]. To analyze PP, we use the method electrodynamic tensor GFs, suitable for metasurfaces with complex configurations, as well as the method of field mode matching. With a linear response, in the case of a conductive layer having a complex three-dimensional structure, it is possible to introduce a volumetric conductivity $\gamma = \sigma/h$ linking the volumetric current density and the electric field $\mathbf{E} = \gamma^{-1}\mathbf{J}$. For example, it can be the bulk specific conductivity of a metal from which a metal film with a two-dimensional structure is made, for example, in the form of holes, slits, protrusions, etc. Two-dimensional periodic (2D-P) structures with periods $d_x, d_y \ll \lambda$ along the *x*, *y* axes are interesting here. For them, an effective surface conductivity tensor $\hat{\sigma}^{ef}$ is introduced by homogenization methods. If the film is dielectric or metal, then the current density of its polarization should be considered as $\mathbf{J}_p = i\omega\varepsilon_0(\varepsilon - 1)\mathbf{E}$, so formally we introduce the volume specific conductivity of the dielectric $\gamma_d(\omega, \mathbf{r}) = i\omega\varepsilon_0(\varepsilon(\mathbf{r}) - 1)$. For metal, the DP in form of Drude-Lorentz $\varepsilon_m = \varepsilon_L - \omega_p^2/(\omega^2 - i\omega\omega_c)$ should be used instead of $\varepsilon$. Similarly, for magnetic properties (for example, a thin ferrite film) we have $\mathbf{J}_m = i\omega\mu_0(\hat{\mu} - \hat{I})\mathbf{H}$, and $\hat{\gamma}_m = i\omega\mu_0(\hat{\mu} - \hat{I})$. Here $\hat{\mu}$ is the magnetic permeability tensor in a constant magnetic field, $\hat{I}$ is unit tensor. Surface magnetic conductivity $\hat{\sigma}_m = \hat{\gamma}_m h$ can also be introduced. It is very interesting to consider the influence of an external magnetic field on the surface conductivity of metal films, graphene sheets and ferrite films, which is also manifested in SD. We are not considering this difficult problem here. We model the metal film by constant Lorentz term $\varepsilon_L \sim 10 - 20$, which runs down to optical frequencies. The formal approach using surface conductivities is also suitable for non-conducting DP films, the polarization current of which is determined by the bias current. The film can lie on the surface of the substrate with DP $\tilde{\varepsilon}$. We consider substrates of both infinite thickness (half-space) and finite thickness *t*, which we assume to be of the order of wavelength or less. In general, the surface conductivity should be understood as the input conductivity of the structure from the vacuum side. Let the $z$ axis be directed perpendicular to the metasurface with the boundary at $z=0$. For 2DEG the current in this direction can be neglected and a surface current density $\mathbf{j}$ can be introduced as $\mathbf{J}(x,y,z) = \mathbf{j}_s(x,y)\delta(z)$, then for a homogeneous continuous film the conductivity is $\sigma = \gamma t$. The surface current density leads to the replacement of the volume integrals of the current density $\mathbf{J}$ by the surface ones $\mathbf{j}$. Surface conductivity is introduced as a limit $\lim(\gamma t)_{t \to 0}$, which for real three-dimensional substances requires an infinitely large volume conductivity. I.e., a real thin layer physically cannot have an arbitrarily small thickness (which is also limited by at least a few atomic layers). For natural two-dimensional substances such as graphene, it can be considered $t=0$ at finite conductivity, although the physical thickness of graphene is 0.34 nm. We do not consider complex metasurfaces with volumetric configurations of the type [47,48] and extended structures of the type [49–51], in which spatial dispersion can be very significant.

**2. Graphene and thin films surface conductivity**

For graphene, within the framework of the Kubo-Greenwood approach, the tensor conductivity has been obtained by integrating over the $\pi^{\pm}$ Brillouin zone for the dispersion in graphene using the Wallace dispersion law obtained in the strong coupling approximation with consideration of the dynamics of $\pi$-electrons:

$$E(\mathbf{p}) = \pm \gamma_0 \sqrt{1 + 4\cos(a_0 p_x)\cos\left(\frac{a_0 p_y}{\sqrt{3}}\right) + 4\cos^2\left(\frac{a_0 p_y}{\sqrt{3}}\right)}. \tag{3}$$

Here, for graphene, $\gamma_0 \approx 3.0$ eV is the overlap integral, $a_0 = 3b_0/(2\hbar)$, $b_0 = 0.142$ nm, $\mathbf{p} = \hbar\mathbf{q}$ is a two-dimensional pulse. In the approximation of the classical kinetic Boltzmann equation (KBE) for the distribution function

$$\partial_t f + e\mathbf{E}\cdot\nabla_{\mathbf{p}} f + \mathbf{v}\cdot\nabla_{\mathbf{r}_\tau} f = J_c(f_{FD}(\mathbf{p}), f(\mathbf{p},\mathbf{r}_\tau,t)). \tag{4}$$

Here $J_c$ is the collision integral. The magnetic field can also be included in (4). The surface current density is determined in Kubo linear approximation [46]

$$\mathbf{j} = \frac{2e}{(2\pi\hbar)^2} \iint_{BZ} \mathbf{v}(\mathbf{p}) f(\mathbf{p}) d^2 p = \hat{\sigma} \mathbf{E}_s. \qquad (5)$$

Here $\mathbf{v}(\mathbf{p}) = \nabla_\mathbf{p} E(\mathbf{p})$, and the integration goes along the Brillouin zone taking into account both types of charge carriers. The calculation of the integral (5) requires the obtaining the solution of the KBE (4), which in the approximation of the relaxation time $J_c = \omega_c [f_{FD}(\mathbf{p}) - f(\mathbf{p}, \mathbf{r}_\tau, t)]$ for the incident plane wave $\mathbf{E} = \mathbf{E}_0 \exp(i[\omega t - k_x x + k_y y])$ is [40]

$$f = f_{FD}(\mathbf{p}) + \delta f = F(\mathbf{p}) + e\left[ \frac{i \nabla_\mathbf{p} f_{FD}(\mathbf{p}) \cdot \mathbf{E}_0 \exp(i[\omega t - \mathbf{k}\mathbf{r}_\tau])}{\omega - \mathbf{k}\cdot\mathbf{v} + i\omega_c} \right]. \qquad (6)$$

Here $\mathbf{k}\mathbf{r}_\tau = k_x x + k_y y$, is the Fermi-Dirac distribution function $f_{FD}(\mathbf{p}) = f_0(\mathbf{p}) = [1 + \exp((E(\mathbf{p}) - \mu_c)/k_B T)]$ with the chemical potential $\mu_c$. The distribution function should be understood as the real part (6). Therefore, from (5), taking into account (3) and (6), one can express the complex conductivity as integral over Brillouin zone:

$$\hat{\sigma}(\omega, \mathbf{k}_\tau, \omega_c) = \frac{2e}{(2\pi\hbar)^2} \int_{BZ} \frac{\nabla_\mathbf{p} f_{FD}(\mathbf{p}) \otimes \mathbf{v}(\mathbf{p})}{\omega - \mathbf{k}\cdot\mathbf{v} + i\omega_c} d^2 p. \qquad (7)$$

In a number of papers (see [40–46]) approximate expressions for (7) are obtained. The conductivity (7) is intraband. The full conductivity of graphene is the sum of the intraband and interband ones: $\hat{\sigma}^{intra}(\omega, \mathbf{k}) = \hat{\sigma}^{intra}(\omega, \mathbf{k}) + \hat{\sigma}^{inter}(\omega, \mathbf{k})$. The intraband one splits into the sum of the electron and hole conductivities $\hat{\sigma}^{intra}(\omega, \mathbf{k}) = \hat{\sigma}^e(\omega, \mathbf{k}) + \hat{\sigma}^h(\omega, \mathbf{k})$. The conductivity with SD reads [40]

$$\sigma_{xx}^{e,h}(\omega, \mathbf{k}) = \frac{\tilde{\sigma}_{xx}^{e,h}(\omega, \mathbf{k}) + k_y \left( \tilde{\sigma}_{xx}^{e,h}(\omega, \mathbf{k}) d_y^{e,h} - \tilde{\sigma}_{yx}^{e,h}(\omega, \mathbf{k}) d_x^{e,h} \right)}{1 + k_x d_x^{e,h} + k_y d_y^{e,h}}.$$

Other components are written out similarly, and

$$\hat{\tilde{\sigma}}^{e,h}(\omega, \mathbf{k}, \mu_c) = \frac{-ie^2}{8k_B T \pi^2} \int_{BZ} \frac{\mathbf{v}^\pm(\mathbf{q}) \otimes \mathbf{v}^\pm(\mathbf{q}) d^2 q}{\cosh^2\left( E^\pm(\mathbf{q}) - \mu_c / 2k_B T \right)(\omega - i\omega_c - \mathbf{k}\mathbf{v}^\pm(\mathbf{q}))},$$

$$\mathbf{d}^{e,h}(\omega, \mathbf{k}, \mu_c) = \frac{-i\omega_c}{\omega F_{e,h}(\omega, \mathbf{k})} \int_{BZ} \frac{\mathbf{v}^\pm(\mathbf{q}) d^2 q}{\cosh^2\left( E^\pm(\mathbf{q}) - \mu_c / 2k_B T \right)(\omega - i\omega_c - \mathbf{k}\mathbf{v}^\pm(\mathbf{q}))}, \qquad (8)$$

$$F_{e,h}(\omega,\mathbf{k},\mu_c) = \int_{BZ} \frac{d^2q}{\cosh^2\left(\mathrm{E}^{\pm}(\mathbf{q}) - \mu_c/2k_BT\right)(\omega - i\omega_c - \mathbf{k}\mathbf{v}^{\pm}(\mathbf{q}))}.$$

Tensor interband conductivity is obtained in [37]:

$$\sigma_{\alpha\beta}(\omega,\mathbf{k}) = \frac{ie^2}{\pi^2\hbar}\left\{\sum_{n=1,2}\int_{BZ}\frac{d^2p\, v_\alpha v_\beta\{f_0(\mathrm{E}_n(\mathbf{p}^-)) - f_0(\mathrm{E}_n(\mathbf{p}^+))\}}{[\mathrm{E}_n(\mathbf{p}^+) - \mathrm{E}_n(\mathbf{p}^-)](\omega\hbar - \mathrm{E}_n(\mathbf{p}^+) + \mathrm{E}_n(\mathbf{p}^-))} \right.$$
$$\left. + 2\omega\hbar\int_{BZ}\frac{d^2p\, v_\alpha^{12} v_\beta^{21}\{f_0(\mathrm{E}_1(\mathbf{p}^-)) - f_0(\mathrm{E}_2(\mathbf{p}^+))\}}{[\mathrm{E}_2(\mathbf{p}^+) - \mathrm{E}_1(\mathbf{p}^-)](\omega^2\hbar^2 - [\mathrm{E}_2(\mathbf{p}^+) - \mathrm{E}_1(\mathbf{p}^-)]^2)}\right\}. \qquad (9)$$

В (9) $\mathrm{E}_n(\mathbf{p}) = \mathrm{E}_n(\mathbf{q}) = (-1)^n \gamma_0 w(\mathbf{q})$, $n=1$

n=1 corresponds to the valence band (VB) $\pi^-$ of the ZB, n=2 corresponds to the conduction band (CB) $\pi^+$, the velocity components are $v_\alpha^{\pm}(\mathbf{q}) = \pm\gamma_0(\partial/\partial q_\alpha)w(\mathbf{q})/\hbar$, and the pulse is $\mathbf{p}^{\pm} = \mathbf{p} \pm \hbar\mathbf{k}/2 = \hbar(\mathbf{q} \pm \mathbf{k}/2)$. The first term in tensor (9) corresponds to the intraband conductivity, and the second term corresponds to the interband conductivity. Tensor (9) acts on the spectral amplitude $E_\beta(\omega,x,y,0)$ of the electric field on graphene: $j_\alpha(\omega,x,y) = \sigma_{\alpha\beta}(\omega,\mathbf{k})E_\beta(\omega,x,y,0)$. The linear term (9) in [37] is obtained based on the representation of the current operator $j_\alpha = -e^2\eta_0\langle\tilde{\psi}^+ m_{\alpha\beta}^{-1}\tilde{\psi}\rangle A_\beta/c$. For the linear dispersion $\mathrm{E} = v_F|\mathbf{p}|$ near the Dirac point, interband conductivity can be represented as scalar

$$\sigma^{\mathrm{inter}}(\omega,\omega_c,k,\mu_c,T) = \frac{-ie^2(\omega-i\omega_c)}{\pi}\int_0^\infty \frac{f_{FD}(-\varepsilon + k\hbar v_F/2) - f_{FD}(\varepsilon - k\hbar v_F/2)}{(\hbar\omega - i\hbar\omega_c \mp k\hbar v_F/2)^2 - 4\varepsilon^2}d\varepsilon. \qquad (10)$$

Here $v_F = \sqrt{3}\gamma_0 a/(2\hbar)$ is the Fermi velocity, $k = \mathbf{k} = \sqrt{k_x^2 + k_y^2}$. For linear dispersion without taking into account SD the intraband conductivity is [39]

$$\sigma^{\mathrm{intra}}(\omega,\omega_c,\mu_c,T) = \frac{ie^2}{\pi\hbar^2(\omega - i\omega_c)}\int_0^\infty (\partial_\varepsilon f_{FD}(\varepsilon) - \partial_\varepsilon f_{FD}(-\varepsilon))\varepsilon d\varepsilon,$$

and the result of integration

$$\sigma^{\mathrm{intra}}(\omega,\omega_c,\mu_c,T) = \frac{k_BT(e^2/\hbar)\varphi(\mu_c,T)}{\pi\hbar\omega_c(1+i\omega/\omega_c)} = \frac{4\sigma_0 k_BT\varphi(\mu_c,T)}{\pi\hbar\omega_c(1+i\omega/\omega_c)}, \qquad (11)$$

$$\varphi(\mu_c,T) = \ln\left(2\cosh\left(\frac{\mu_c}{2k_BT}\right)\right). \qquad (12)$$

Simple approximation without SD ($k=0$) and for $k_BT \ll \mu_c$, $k_BT \ll \hbar\omega$ leads [39]

$$\sigma^{\text{inter}}(\omega,\omega_c,\mu_c) \approx \frac{-i\sigma_0}{\pi} \ln\left(\frac{2|\mu_c| - (\omega - i\omega_c)\hbar}{2|\mu_c| + (\omega - i\omega_c)\hbar}\right). \tag{13}$$

Here $\sigma_0 = e^2/(4\hbar)$, $\xi_0 = \sigma_0 \eta_0 = \pi\alpha_0 = 0.0229$, $\eta_0 = \sqrt{\mu_0/\varepsilon_0}$, $\alpha_0 = 1/137$ is the fine structure constant. Next, we will use normalized conductivities. Formula (13) is not exact, so for $T \neq 0$ and k=0 (without SD) we have got a more accurate formula

$$\sigma^{\text{inter}}(\omega,\omega_c,T) = \sigma_0 + \frac{i\sigma_0}{\pi}\left[\frac{\beta\exp(-\alpha)}{\beta^2 - 4\alpha^2}\left(1 + \frac{3\alpha+1}{\beta^2 - 4\alpha^2} + \frac{2\alpha^2}{(\beta^2 - 4\alpha^2)^2} - \frac{\exp(-2\alpha)}{2}\right) + \right.$$
$$\left. + \frac{1/4}{1+\exp(\alpha/2)}\ln\left(\frac{\beta^2 + \alpha\beta - 2\alpha^2}{\beta^2 - \alpha\beta - 2\alpha^2}\right) + \frac{1/4}{1+\exp(-\alpha/2)}\ln\left(\frac{\beta-\alpha}{\beta+\alpha}\right)\right]. \tag{14}$$

Here $\alpha = \mu_c/(k_B T)$ and $\beta = \hbar(\omega - i\omega_c)/(2k_B T)$. Also from (10) we have obtained an approximate formula for small SD ($k \neq 0$). We have used in calculation the results [40]

$$\sigma_{xx}(\omega,\omega_c,\mathbf{k}) = \sigma^{\text{intra}}\left[1 + \frac{v_F^2}{4(\omega - i\omega_c)^2}\left(3 - \frac{2i}{\omega/\omega_c}\right)k_x^2 + \frac{v_F^2}{4(\omega - i\omega_c)^2}k_y^2\right] + \sigma^{\text{inter}}(k), \tag{15}$$

$$\sigma_{xy}(\omega,\omega_c,\mathbf{k}) = \sigma_{\text{intra}}\left[\frac{v_F^2}{2(\omega - i\omega_c)^2}k_x k_y\right], \tag{16}$$

obtained in the Bhatnagar-Gross-Crook approximation. For a strong SD, we have used numerical calculation of integrals.

Let estimate the scalar conductivity $\sigma$ for continuous thin metal films. Consider copper, aluminum and silver layers with thicknesses of tens of nanometers at a temperature of $T = 300$ K. If the thickness of the skin layer $\delta$ of a non-magnetic metal is much less than the wavelength $\lambda$, thickness $h$ and greater than the free path length $\lambda_0$, then $\sigma(\omega) = 2\gamma\delta/(1+i) = (1-i)\sqrt{2\gamma/(\omega\mu_0)}$. Here multiplier 2 means that both film boundaries work in parallel. The depth of penetration into silver and other metals is always more than 200 nm [52]. Therefore, for thicknesses $h$ of the order of hundreds of nanometers or more, this model is not applicable or is only limited. In this case, refined models should be used. Since the film works as a shunt, its input conductivity is $\sigma(\omega) = i\rho^{-1}\tan(k_z h)$. For the E-wave, there is a normalized wave impedance $\rho = k_z/(\omega\varepsilon_0\varepsilon)$, therefore, with a small thickness $\sigma(\omega) = ihk_0\varepsilon/\eta_0$, and for the H-wave $\rho = \omega\mu_0/k_z$, and respectively $\sigma(\omega) = ih(k_0^2\varepsilon - k_x^2)/(k_0\eta_0)$. Usually in metal $|\varepsilon| \gg 1$, so both values coincide and determine the usual surface conductivity. For the surface density in a metal sheet, the surface concentration of electrons should be assumed $n_S = Nh$, where the volume densities for copper $N = 8.45 \cdot 10^{28}$ m$^{-3}$, for aluminum $N = 18.18 \cdot 10^{28}$ m$^{-3}$, and for silver $N = 5 \cdot 10^{28}$ m$^{-3}$. The effective masses of charge carriers are approximately equal to the mass of an electron. Here the thickness of the layer *h* should be entered. The model works well when the thickness does not exceed twice the thickness of the skin layer and $h > \lambda_0$. Otherwise, the phase incursion can no longer be neglected, and it is necessary to use two-way boundary conditions or consideration of the layer based on the transmission matrix. In the case $h < \lambda_0$, the nature of the transverse electron transport is ballistic, and the current depends on the density of states (the number of conduction modes). In this case, according to Landauer's concept, the number of conduction modes can be estimated approximately as the number of de Broglie half-waves that fit on the size *h*, i.e. the conductivity is no longer proportional to *h*, but depends on this size in a complicated stepwise manner. Consider the correction term to the Drude formula for the metal layer. This formula $\hat{\sigma}(\omega) = \hat{\sigma}_0 / (1 + i\omega/\omega_c)$, corresponds to the DP of the Drude-Lorentz metal at $\varepsilon_L = \varepsilon = 1$. If we consider a sheet of metal in the medium, then $\gamma = i\omega\varepsilon_0(\varepsilon - \tilde{\varepsilon})$. Therefore, in the general case, a term $i\omega\varepsilon_0(\varepsilon_L - \tilde{\varepsilon})h$ should be added to conductivity. In particular, it leads to a downward shift of the resonant frequency of the PP over the metal half-space (Zenneck PP).

Let's consider a model of the conductivity of a metal film with a more accurate account of its thickness. Let the film be placed in an infinite medium with DP $\tilde{\varepsilon}$. The input impedance at the left edge of the film is equal to $Z_{in} = Z_M(\eta_0\rho + iZ_M \tan(\theta))/(Z_M + i\eta_0\rho \tan(\theta))$. Here $\rho$ is the normalized wave resistance of the medium for *p*- or *s*-polarization $Z_M$ is the wave resistance of the metal, $\theta = h\sqrt{k_0^2\varepsilon - k_x^2}$. The value $|\varepsilon|$ is usually very large for metal (except for the ultraviolet region), therefore $\theta \approx hk_0\sqrt{\varepsilon}$, as well as $Z_M \approx \eta_0/\sqrt{\varepsilon}$ for both polarizations. With a small thickness the film is equivalent to the surface resistance $Z_S$, at the same time $Z_{in} = [Z_S^{-1} + (\eta_0\rho)^{-1}]^{-1}$, from where $\sigma = Z_{in}^{-1} + (\eta_0\rho)^{-1}$. Such conductivity depends on the polarization and the propagation angle (i.e. from $k_x$). Here is its value for a very small thickness and *p*-polarization:

$$\sigma\eta_0 = \sqrt{\tilde{\varepsilon}}\left[\left(1+ih\sqrt{1-n^2}k_0\varepsilon\sqrt{\tilde{\varepsilon}}\right)/\left(\sqrt{1-n^2}+ihk_0\sqrt{\tilde{\varepsilon}}\right)-1/\sqrt{1-n^2}\right].$$

Here $n = k_x/(k_0\sqrt{\tilde{\varepsilon}})$ is the deceleration factor. In general, the film should be described by a transmission matrix, and the results given will be obtained if the phase gain and the impedance of the film are small. By changing *h* within the specified limits, it is possible to significantly change the surface conductivity. Since the volumetric specific conductivity $\gamma$ is related to the free path length $\lambda_0$ and the Fermi velocity $v_F$ by the ratio $\gamma = e^2 N\lambda_0/(m_e v_F)$, then for copper with conductivity $\gamma = 5.81\cdot 10^7$ Cm/m we have $\lambda_0 = 38.9$ nm, while $\omega_p = 2.2\cdot 10^{16}$, $\omega_c = 5\cdot 10^{13}$ Hz. For aluminum $\gamma = 3.7\cdot 10^7$ Cm/m, $\omega_p = 3.23\cdot 10^{16}$, $\omega_c = 1.6\cdot 10^{14}$ Hz. For silver, $\gamma = 6.25\cdot 10^7$ Cm/m, $\omega_p = 1.57\cdot 10^{16}$, $\omega_c = 3.5610^{13}$ Hz. The $\lambda_0$ for aluminum and silver are equal to 14.94 и 59.42 nm, respectively. If the value of *h* is significantly larger $\lambda_0$, then in such a layer you can use this volumetric value of $\lambda_0$ the chipboard and take $\sigma_0 = \gamma h$. The effective two-dimensional chipboard changes with decreasing thickness, which is fundamentally limited to several atomic layers. The conductivity of 2-DEG then requires quantum consideration and depends on the adjacent layers and a number of other factors. It is necessary to solve quantum-dimensional problems and determine the number of conduction modes. It is possible to estimate the static conductivity approximately as $\sigma_0 = \gamma h$, since our goal is not to calculate the specified value, depending on many factors, but to obtain the dispersion of PP. A comparison of theoretical and experimental results for such dispersion makes it possible to determine the dynamic conductivity $\sigma_0$ as a function of frequency.

## 3. Dispersion equations

Let consider a metasurface at $z=0$ in the form of conducting infinitely long along $x$-axis lonely strips of width $w$ with surface conductivity $\hat{\sigma}$ or periodically located along the $y$-axis such strips with period $b$ and directed along the $x$-axis. We also consider infinite width $w \to \infty$ a limiting case. A surface current density is $\mathbf{j}(x,y) = (\mathbf{x}_0 j_x(y) + \mathbf{y}_0 j_y(y))\exp(-ik_x x)$, $|y| < w/2$, that satisfies the condition $j_x(y) = \sigma_{xx} E_x(y) + \sigma_{xy} E_y(y)$, $j_y(y) = \sigma_{yx} E_x(y) + \sigma_{yy} E_y(y)$, $\sigma_{xy} = \sigma_{yx}$. The problem for film in medium with DP $\tilde{\varepsilon}$ is solved using the only electrical vector potential

$$\mathbf{A}(\mathbf{r}) = \mathbf{x}_0 \int_{-\infty}^{\infty}\int_{-w/2}^{w/2} \tilde{G}_\omega(\mathbf{r}, x', y', 0) j_x(x', y') dx' dy' + \mathbf{y}_0 \int_{-\infty}^{\infty}\int_{-w/2}^{w/2} \tilde{G}_\omega(\mathbf{r}, x', y', 0) j_x(x', y') dx' dy'. \quad (17)$$

For periodic structure the Green's function (GF) $\tilde{G}_\omega$ is

$$\tilde{G}_\omega(\mathbf{r},\mathbf{r}') = \frac{1}{4\pi^2 b} \sum_{n=-\infty}^{\infty}\int_{-\infty}^{\infty} dk_x \int_{-\infty}^{\infty} dk_z \frac{\exp(-ik_x(x-x') - i\tilde{k}_{yn}(y-y') - ik_z(z-z'))}{k_x^2 + \tilde{k}_{yn}^2 + k_z^2 - k_0^2 \tilde{\varepsilon}}.$$

or

$$\tilde{G}_\omega(\mathbf{r},\mathbf{r}') = \frac{1}{2\pi b} \sum_{n=-\infty}^{\infty}\int_{-\infty}^{\infty} dk_x \frac{\exp(-ik_x(x-x') - i\tilde{k}_{yn}(y-y') \mp i\tilde{k}_{zn}|z-z'|)}{2i\tilde{k}_{zn}}. \quad (18)$$

Here $\tilde{k}_{zn} = \sqrt{k_0^2 \tilde{\varepsilon} - k_x^2 + \tilde{k}_{yn}^2}$, $\tilde{k}_{yn} = k_y + (2n\pi/b)$. We can consider plasmons with arbitrary $k_y$, in particular, $k_y = 0$. For non periodic structure $k_z = \sqrt{k_0^2 \tilde{\varepsilon} - k_x^2 + k_y^2}$ and

$$G_\omega(\mathbf{r},\mathbf{r}') = \frac{1}{(2\pi)^2} \int_{-\infty}^{\infty}\int_{-\infty}^{\infty} dk_y dk_x \frac{\exp(-ik_x(x-x') - ik_y(y-y') \mp ik_z|z-z'|)}{2ik_z}. \quad (19)$$

The GFs (18), (19) are the Fourier transform of the function $g_{\omega n}(\mathbf{k}, z) = -i\exp(-i\tilde{k}_{zn}|z|)/(2\tilde{k}_{zn})$ and $g_\omega(\mathbf{k}, z) = -i\exp(-i\tilde{k}_z|z|)/(2\tilde{k}_z)$. To determine

the field on the surface we will put $z = z' = 0$. For this $g_\omega(\mathbf{k},0) = g_\omega(\mathbf{k}) = -i/\tilde{k}_z$.

Acting on (1) by the operator $(i\omega\varepsilon_0\varepsilon)^{-1}(k_0^2\varepsilon + \nabla\nabla\cdot)$, we obtain the equation

$$\mathbf{E}(\mathbf{r}) = \int_{-\infty}^{\infty}\int_{-w/2}^{w/2} \hat{G}_\omega(\mathbf{r}, x', y', 0)\mathbf{j}(x', y')dx'dy' . \tag{20}$$

We will be interested in the tangential electric field $\mathbf{E}_\tau(x, y)$ at $z = 0$, for which, as for current density and GF (19), we will consider the Fourier transform ($\alpha = x, y$)

$$\hat{G}_\omega(x, y) = \frac{1}{(2\pi)^2} \int_{-\infty}^{\infty}\int_{-\infty}^{\infty} \hat{g}(k_x, k_x)\exp(-ik_x x - ik_y y)dk_x dk_y ,$$

$$E_\alpha(x, y) = \frac{1}{(2\pi)^2} \int_{-\infty}^{\infty}\int_{-\infty}^{\infty} e_\alpha(k_x, k_y)\exp(-ik_x x - ik_y y)dk_x dk_y ,$$

$$j_\alpha(x, y) = \frac{1}{(2\pi)^2} \int_{-\infty}^{\infty}\int_{-w/2}^{w/2} j_\alpha(k_x, k_y)\exp(-ik_x x - ik_y y)dk_x dk_y ,$$

$$\hat{g}(k_x, k_x) = \frac{-1}{2k_0\sqrt{\tilde{\varepsilon}}\tilde{k}_z}\begin{bmatrix} k_0^2\tilde{\varepsilon} - k_x^2 & -k_x k_y \\ -k_x k_y & k_0^2\tilde{\varepsilon} - k_y^2 \end{bmatrix}, \tag{21}$$

$$\begin{pmatrix} e_x(k_x, k_y) \\ e_y(k_x, k_y) \end{pmatrix} = \hat{g}(k_x, k_x)\begin{pmatrix} j_x(k_x, k_y) \\ j_y(k_x, k_y) \end{pmatrix}.$$

These relations can be generalized to the current density on a substrate of thickness $t$ c DP $\varepsilon$ in vacuum ($\tilde{\varepsilon} = 1$):

$$\hat{g}(k_x, k_x) = \frac{i\eta_0}{\kappa^2}\begin{bmatrix} \dfrac{k_x^2}{k_0 d^e} - \dfrac{k_y^2 k_0}{d^m} & -k_x k_y\left(\dfrac{1}{k_0 d^e} + \dfrac{k_0}{d^m}\right) \\ -k_x k_y\left(\dfrac{1}{k_0 d^e} + \dfrac{k_0}{d^m}\right) & \dfrac{k_y^2}{k_0 d^e} - \dfrac{k_x^2 k_0}{d^m} \end{bmatrix}. \tag{22}$$

Here $d^e = (1/\kappa_1 + 1/Z^e)$, $d^m = (\kappa_1 + 1/Z^m)$, $\kappa_1 = \sqrt{\kappa^2 - k_0^2}$, $\kappa_2 = \sqrt{\kappa^2 - k_0^2\varepsilon}$, and also the values are indicated

$$Z^e = \frac{\kappa_2}{\varepsilon}\frac{\varepsilon\kappa_1 + \kappa_2\tanh(\kappa_2 t)}{\kappa_2 + \varepsilon\kappa_1\tanh(\kappa_2 t)}, \quad Z^m = \frac{1}{\kappa_2}\frac{\kappa_2 + \kappa_1\tanh(\kappa_2 t)}{\kappa_1 + \kappa_2\tanh(\kappa_2 t)}.$$

For the case of a half-space $d^e = (1/\kappa_1 + \varepsilon/\kappa_1)$, $d^m = (\kappa_1 + \kappa_2)$. For the case of an electric wall at z=−t we have $d^e = (1/\kappa_1 + \varepsilon\coth(\kappa_2 t)/\kappa_2)$, $d^m = (\kappa_1 + \kappa_2\coth(\kappa_2 t))$. In the case of t=0, or for $\varepsilon = 1$ the tensor (22) passes to (21).

For thin perforated films in vacuum, when there is a period comparable to the wavelength, an approximate admittance can be introduced on the entire surface $\sigma_{xx} = \sigma w/a$, $\sigma_{yy} = 0$. In particular, this applies to hole graphene [53]. For a metasurface of 2D-periodically arranged long strips of length $l$ and width $w$ ($l \gg w$), it will be

$$\mathbf{A}(\mathbf{r}) = \mathbf{x}_0 \int_{-l/2}^{l/2} \int_{-w/2}^{w/2} \tilde{G}(\mathbf{r},\mathbf{r}') j_x(x',y') dx' dy'$$

with 2D-periodical scalar GF

$$\tilde{G}(\mathbf{r},\mathbf{r}',\omega) = \frac{1}{2\pi ab} \sum_{m=-\infty}^{\infty} \sum_{m=-\infty}^{\infty} \int_{-\infty}^{\infty} \frac{\exp(-ik_{xm}(x-x') - i\tilde{k}_{yn}^2(y-y') - ik_z(z-z'))}{\tilde{k}_{xm}^2 + \tilde{k}_{yn}^2 + k_z^2 - k_0^2 \tilde{\varepsilon}} dz. \quad (23)$$

Here $\tilde{k}_{xm} = 2m\pi/a_x + k_x$, $\tilde{k}_{yn} = 2n\pi/a_y + k_y$. The solution of the boundary value problem is reduced to the DE, from which the dispersion of PP is determined. For a model of the conductivity of a film with holes, integral equations should be constructed with respect to the electric field on the holes. However, with very small perforations, the theory of leakage can be used. If the holes are elliptical or rectangular forms with orientation along one of the directions, then the surface conductivity tensor can be obtaining based on M. Garnett type formulas with the introduction of depolarization coefficients.

In this paper, we consider thin 2DEGS, i.e. we neglect the transverse current. Below we will consider the effect of the film thickness $h$. For a metal film with DP, $\varepsilon(\omega) = \varepsilon_L - \omega_p^2/(\omega^2 - i\omega\omega_{co})$ we have conductivity $\sigma(\omega) = i\omega\varepsilon_0[\varepsilon_L - \omega_p^2/(\omega^2 - i\omega\omega_{co})]h$, i.e. $j_\alpha(x,y) = J_\alpha(x,y,z_n)h$, or formally, the volumetric current density can be written as $J_\alpha(x,y,z_0) = j_\alpha(x,y)\delta(z-z_0)$. For a delayed plasmon at $z=0$ and even in $z$ with low dissipation, we have $j_\alpha(z) = j_{0\alpha}\cosh(i\chi z) \approx j_{0\alpha}\cosh(\alpha z)$, $\chi = \chi' - i\chi''$. The attenuation constant has the form

$$\chi'' = k_0\sqrt{\frac{\sqrt{[\varepsilon_L(\omega^2 + \omega_{co}^2) - \omega_p^2]^2 + \omega_p^4\omega_{co}^2/\omega^2} - [\varepsilon_L(\omega^2 + \omega_{co}^2) - \omega_p^2]}{2(\omega^2 + \omega_{co}^2)}},$$

and $\cosh(\chi''h/2)$ gives a weakening of the current in the center compared to the surface. Obviously, the thickness can always be reduced so that the current and field are constant across the section. The value $\varepsilon_L$ corresponds to the contribution to DP from lattice polarization and interband transitions. The relation of specific and surface conductivities is given by the relation $\gamma = \sigma/h$.

Let's denote the 2D vector $\boldsymbol{\rho} = \mathbf{x}_0 x + \mathbf{y}_0 y$. We introduce the dyadic GF $\hat{G}^{ee}(\mathbf{r}) = (-ik_0\hat{I} + ik_0^{-1}\nabla\otimes\nabla)\eta_0 G(\mathbf{r})$, defining the electric field $\mathbf{E}(\mathbf{r}) = (ik_0\sqrt{\tilde{\varepsilon}})^{-1}\eta_0(\nabla\otimes\nabla + \hat{I}k_0\sqrt{\tilde{\varepsilon}})\mathbf{A}(\mathbf{r})$ and impose boundary conditions:

$$j_{n\alpha}(x,y) = \sigma_{\alpha\beta}E_\beta(x,y,z_n). \tag{24}$$

Here we have coupled sheets of graphene for certainty: $n=2$, $z_1=0$, $z_2=d$. It is convenient to enter the function $\psi = \exp(-\kappa_z d/2)$, $\kappa_z = \sqrt{k_x^2 + k_y^2 - \tilde{\varepsilon}k_0^2} = ik_z$ and dimensionless (normalized) conductivity: $\hat{\xi} = \hat{\sigma}\eta_0$. The representation of fields in terms of GF leads to Lippmann-Schwinger type equations. Turning to the spatial spectrum, we obtain the first equation in (24 which reads

$$[1 - \xi_{xx}g_{xx}(\mathbf{k}) - \xi_{xy}g_{xy}(\mathbf{k})]j_{1x}(\mathbf{k}) - [\xi_{xx}g_{xy}(\mathbf{k}) + \xi_{xy}g_{yy}(\mathbf{k})]j_{1y}(\mathbf{k}) - \\ -\psi^2[\xi_{xx}g_{xx}(\mathbf{k}) + \xi_{xy}g_{xy}(\mathbf{k})]j_{2x}(\mathbf{k}) - \psi^2[\xi_{xx}g_{xy}(\mathbf{k}) + \xi_{xy}g_{yy}(\mathbf{k})]j_{2y}(\mathbf{k}) = 0 \tag{25}$$

To get the second equation, you need to make substitutions $x \leftrightarrow y$. To get the third equation, it is necessary to make substitutions $1 \leftrightarrow 2$ in the first one. The fourth equation can be obtained by corresponding substitutions from the first, or from the second. To get it from the second equation, we make the replacement $1 \leftrightarrow 2$. In this case the DE takes the form

$$\det\begin{bmatrix} \hat{I} - \hat{A}(\mathbf{k}) & -\psi^2(k)\hat{A}(\mathbf{k}) \\ -\psi^2(k)\hat{A}(\mathbf{k}) & \hat{I} - \hat{A}(\mathbf{k}) \end{bmatrix} = 0, \tag{26}$$

$$\hat{A}(\mathbf{k}) = \begin{bmatrix} \xi_{xx}g_{xx} + \xi_{xy}g_{xy} & \xi_{xx}g_{xy} + \xi_{xy}g_{yy} \\ \xi_{xx}g_{xy} + \xi_{xy}g_{yy} & \xi_{yy}g_{yy} + \xi_{xy}g_{xy} \end{bmatrix}. \qquad (27)$$

The determinant of the block matrix in (26) is equal to $\det^2(\hat{I} - \hat{A}(\mathbf{k})) - \psi^4(k)\det^2(\hat{A}(\mathbf{k}))$. The matrix (27) has the form $A_{11} = a(\mathbf{k})$, $A_{12} = A_{21} = b(\mathbf{k})$, $A_{22} = c(\mathbf{k})$, therefore, we have DE

$$[(1-a(\mathbf{k}))(1-c(\mathbf{k})) - b^2(\mathbf{k})] = \pm\psi^2(k)(a(\mathbf{k})c(\mathbf{k}) - b^2(\mathbf{k})). \qquad (28)$$

It defines two equations. If we are interested in very slow PP, then with a sufficiently large distance $d$ between the sheets $|\psi^2(k)| = |\exp(-\kappa_z d)| \ll 1$. Assuming $\psi^2(k) = 0$, we see that the two branches of DE (28) merge into one:

$$(1-a(\mathbf{k}))(1-c(\mathbf{k})) - b^2(\mathbf{k}) = 0. \qquad (29)$$

It splits into two independent ones if $b(\mathbf{k}) = 0$:

$$1 - a(\mathbf{k}) = 0, \qquad (30)$$

$$1 - c(\mathbf{k}) = 0. \qquad (31)$$

Consider the possibility of obtaining them. Let the plasmon go along one of the axes. Then $g_{xy} = 0$. Let this be the $x$-axis for certainty. Next, we always choose the $x$-axis so that it does not limit the generality. Then $k_y = 0$, $g_{xx} = -k_z/(2k_0) = i\sqrt{n^2 - 1}/2$, $g_{yy} = -k_0/(2k_z) = -i/(2\sqrt{n^2 - 1})$, while $a(\mathbf{k}) = \xi_{xx}g_{xx}$, $b(\mathbf{k}) = \xi_{xy}g_{yy}$, $c(\mathbf{k}) = \xi_{yy}g_{yy}$. We have introduced a complex deceleration $n = n' - in'' = k_z/k_0$. From DE (29) we have $(1 - \xi_{xx}(\mathbf{k})g_{xx}(\mathbf{k}))(1 - \xi_{yy}(\mathbf{k})g_{yy}(\mathbf{k})) = \xi_{xy}^2(\mathbf{k})g_{yy}^2$. This equation decays into (30) and (31) if the conduction tensor is reduced to the main axes, i.e. if $\xi_{xy} = 0$. We get,

$1 - \xi_{xx}(\mathbf{k})g_{xx}(\mathbf{k}) = 0$, $1 - \xi_{yy}(\mathbf{k})g_{yy}(\mathbf{k}) = 0$, or $\xi_{xx}(\mathbf{k})k_z(k) = 2k_0\sqrt{\tilde{\varepsilon}}$ and $\xi_{yy}(\mathbf{k})k_0\sqrt{\tilde{\varepsilon}} = 2k_z(k)$. Squaring them, we have $k_x^2 = k_0^2\tilde{\varepsilon}(1 - 4/\xi_{xx}^2(\mathbf{k}))$, $k_x^2 = k_0^2\tilde{\varepsilon}(1 - \xi_{xx}^2(\mathbf{k})/4)$. Extracting square roots, we get two solutions

$$k_x/k_0 = n_x = \pm\sqrt{\tilde{\varepsilon} - 4\tilde{\varepsilon}/\xi_{xx}^2(\omega, \mathbf{k}, T, \mu_c)}, \qquad (32)$$

$$k_x/k_0 = n_x = \pm\sqrt{\tilde{\varepsilon} - \xi_{yy}^2(\omega, \mathbf{k}, T, \mu_c)/(4\tilde{\varepsilon})}. \qquad (33)$$

The above equations were previously obtained by field mode matching in [36] without SD taking onto account. In fact, with SD, these are implicit equations. Roots with different signs give waves in two directions.

Let consider the single thin metal film ($n=1$) located in a medium with DP $\tilde{\varepsilon}$. Using the spatial-spectral representation for the scalar GF, we find for the electric vector potential $\mathbf{A}(\mathbf{r}) = \mathbf{x}_0 A_x(\mathbf{r}) + \mathbf{y}_0 A_y(\mathbf{r})$: $\mathbf{A}(\mathbf{r}) = (2ik_z)^{-1} \exp(-i\boldsymbol{\rho}\mathbf{k} - ik_z|z|)\mathbf{j}_1(\mathbf{k})$, as well as an electric field $\mathbf{E}(\mathbf{r}) = \mathbf{E}_0(\mathbf{k})\exp(-i\mathbf{k}\boldsymbol{\rho} - ik_z|z|)$:

$$E_{0(x,y)}(\mathbf{k}) = -\frac{(k_0^2\tilde{\varepsilon} - k^2)j_{1(x,y)}(\mathbf{k}) - k_x k_y j_{1(y,x)}(\mathbf{k})}{2\omega\varepsilon_0\tilde{\varepsilon}k_z}, \tag{34}$$

$$E_{0z}(\mathbf{k}) = \text{sgn}(z)\frac{k_x j_{1x}(\mathbf{k}) + k_y j_{1y}(\mathbf{k})}{2\omega\varepsilon_0\tilde{\varepsilon}}. \tag{35}$$

We will be interested in surface waves for which $|\mathbf{k}^2| = |k_x^2 + k_y^2| > k_0^2\tilde{\varepsilon}$. For such waves, there are the magnetic field components:

$$H_x(\mathbf{r}) = \text{sgn}(z)j_{1y}(\mathbf{k})\text{sgn}(z)\exp(-i\mathbf{k}\boldsymbol{\rho} - ik_z|z|)/2,$$

$$H_y(\mathbf{r}) = -\text{sgn}(z)j_{1x}(\mathbf{k})\text{sgn}(z)\frac{\exp(-i\mathbf{k}\boldsymbol{\rho} - ik_z|z|)}{2},$$

$$H_z(\mathbf{r}) = (k_y j_{1x}(\mathbf{k}) - k_x j_{1y}(\mathbf{k}))\frac{\exp(-i\mathbf{k}\boldsymbol{\rho} - ik_z|z|)}{2k_z}.$$

It follows from the first two equations that the components of the tangential magnetic field have jumps equal to their transverse components: $\mathbf{j}_1 = \mathbf{z}_0 \times (\mathbf{H}^+ - \mathbf{H}^-)$. By virtue of the continuity equation, the magnitude $\rho_\tau(\mathbf{k}) = [k_x j_{0x}(\mathbf{k}) + k_y j_{0y}(\mathbf{k})]/\omega$ is the spatial-spectral density of the surface charge, therefore the normal component of the electric field has the jump $E_{0z}(\mathbf{r}_\tau, z+0) - E_{0z}(\mathbf{r}_\tau, z-0) = \rho_\tau(\mathbf{r}_\tau)/(\varepsilon_0\tilde{\varepsilon})$. In order to obtain a DE, it is enough to require the fulfillment of the generalized Ohm's law: $j_{1x} = J_{0x}/h = \sigma_{xx}(\omega)E_{0x} + \sigma_{xy}(\omega)E_{0y}$, $j_{0y} = J_{0y}/h = \sigma_{yx}(\omega)E_{0x} + \sigma_{yy}(\omega)E_{0y}$. In this case, we assume the fulfillment of the Onzager-Casimir relations $\sigma_{yx} = \sigma_{xy}$. For a solid metal film we have $\sigma_{yy} = \sigma_{xx} = \sigma$, $\sigma_{xy} = 0$. Tensor conductivity can be introduced for a conductive film with conductivity $\sigma$ if periodic holes are made on it. Let the film be perforated with periodic rectangular holes $l \times w$ with period $a = b$ of a size much smaller than the wavelength. Then we can introduce an approximate effective conductivity $\sigma_{xx} = \sigma(1 - w^2/a^2)$, $\sigma_{yy} = \sigma(1 - l^2/a^2)$. The film in an medium with DP $\tilde{\varepsilon}$ is described by an electric current of polarization $\mathbf{J}_p(\omega) = i\omega\varepsilon_0(\varepsilon(\omega) - \tilde{\varepsilon})\mathbf{E}(\omega)$. Here there is the magnetic wall in the center of film. On the other hand, we have $\varepsilon(\omega) = \tilde{\varepsilon} - i\gamma(\omega)/(\varepsilon_0\omega)$, from where we get the generalized conductivity $\gamma(\omega) = i\varepsilon_0\omega[\varepsilon(\omega) - \tilde{\varepsilon}] = i\varepsilon_0\omega(\varepsilon_L - \tilde{\varepsilon}) - i\varepsilon_0\omega_p^2/(\omega - i\omega_c)$. We have the equations $j_{1x} = \sigma E_{0x}$, $j_{1y} = \sigma E_{0y}$, or

$$j_{1x}\left[1 + \sigma\eta_0 \frac{k_0^2\tilde{\varepsilon} - k_x^2}{2k_0\tilde{\varepsilon}k_z}\right] - j_{1y}\frac{\sigma\eta_0 k_x k_y}{2k_0\tilde{\varepsilon}k_z} = 0, \tag{36}$$

$$-j_{1x}\frac{\sigma\eta_0 k_x k_y}{2k_0\tilde{\varepsilon}k_z} + j_{1y}\left[1 + \sigma\eta_0 \frac{k_0^2\tilde{\varepsilon} - k_y^2}{2k_0\tilde{\varepsilon}k_z}\right] = 0. \tag{37}$$

The general DE follows from (36), (37)

$$\left[1 + \sigma\eta_0 \frac{k_0^2\tilde{\varepsilon} - k_x^2}{2k_0\tilde{\varepsilon}k_z}\right]\left[1 + \sigma\eta_0 \frac{k_0^2\tilde{\varepsilon} - k_y^2}{2k_0\tilde{\varepsilon}k_z}\right] - \frac{(\sigma\eta_0 k_x k_y)^2}{4k_0^2\tilde{\varepsilon}^2 k_z^2} = 0. \tag{38}$$

It describes the surface 2D PP along metal film, described by surface conductivity. A more rigorous DE is obtained based on the mode matching method [54]. At a given frequency we have complex solutions in the form $k_x(\omega)$ and $k_y(\omega)$. By setting the direction of the wave along one of the axes, for example, by putting $k_y = 0$, we obtain the known equations

$$1 + \frac{\sigma(\omega)\eta_0\sqrt{k_0^2\tilde{\varepsilon} - k_x^2}}{2k_0\tilde{\varepsilon}} = 0, \tag{39}$$

$$1 + \frac{\sigma(\omega)\eta_0 k_0\sqrt{\tilde{\varepsilon}}}{2\sqrt{k_0^2\tilde{\varepsilon} - k_x^2}} = 0, \tag{40}$$

which correspond to (32), (33). For metallic film they have the forms $k_x = k_0\sqrt{\tilde{\varepsilon} + 4/(k_0 h(\varepsilon/\tilde{\varepsilon} - 1))^2}$, $k_x = k_0\sqrt{\tilde{\varepsilon} + (k_0 h(\varepsilon - \tilde{\varepsilon}))^2/4}$. If $|\varepsilon| \gg \tilde{\varepsilon}$ that will be $k_x = k_0\sqrt{\tilde{\varepsilon} + 4\tilde{\varepsilon}^2/(k_0 h\varepsilon)^2}$, $k_x = k_0\sqrt{\tilde{\varepsilon} + (k_0 h\varepsilon/2)^2}$. Rigorous E-PP analysis leads to DE [55]

$$k_x^2 = k_0^2 \tilde{\varepsilon} \frac{\varepsilon(\tilde{\varepsilon} - \varepsilon \tanh^2(\vartheta))}{\tilde{\varepsilon}^2 - \varepsilon^2 \tanh^2(\vartheta)} \quad k_x^2 = k_0^2 \tilde{\varepsilon} \frac{\varepsilon(\tilde{\varepsilon} \tanh^2(\vartheta) - \varepsilon)}{\tilde{\varepsilon}^2 \tanh^2(\vartheta) - \varepsilon^2},$$

respectively for the magnetic and electric walls in the center of the film. Here $\vartheta = h\sqrt{k_x^2 - k_0^2 \tilde{\varepsilon}}/2$. Slow PP occurs when $\tilde{\varepsilon}^2 \approx \varepsilon^2 \tanh^{\pm 2}(\vartheta)$. Considering the thickness is very small, we have $\tilde{\varepsilon}^2 \approx \varepsilon^2 \vartheta^{\pm 2}$, from where $k_x = k_0 \sqrt{\tilde{\varepsilon} + 4\tilde{\varepsilon}^2/(k_0 h \varepsilon)^2}$, $k_x = k_0 \sqrt{\tilde{\varepsilon} + (k_0 h \varepsilon/2)^2}$, that is an approximate approach based on surface conductivity corresponds to the mode-matching method.

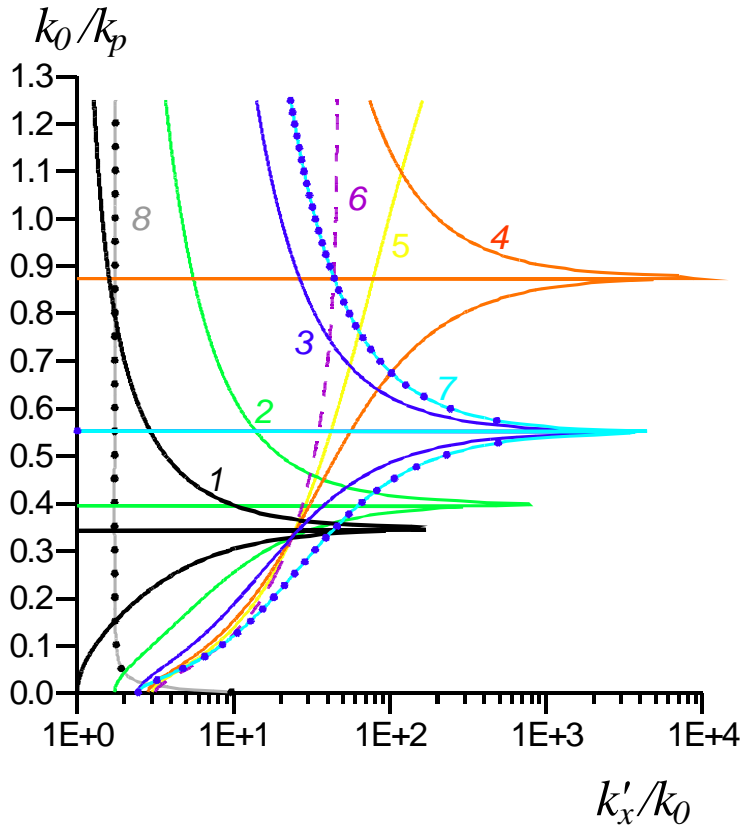

Figure 1. Normalized dispersion for E-PP (curves 1–7) and H-PP (8) along silver DEG ($\varepsilon_L = 9.3$, $\omega_p = 1.57 \cdot 10^{16}$, $\omega_{co} = 3.56 \cdot 10^{13}$) at $h$=5 nm (curves 1–6) и $h$=3 nm (7,8) for $\tilde{\varepsilon} = 1$ (curve 1), $\tilde{\varepsilon} = 3$ (2,8), $\tilde{\varepsilon} = 6$ (3,7), $\tilde{\varepsilon} = 8$ (4), $\tilde{\varepsilon} = 9$ (5), $\tilde{\varepsilon} = 10$ (6) (frequencies in Hz)

The wave corresponding to the branch (39) has no components $E_x$ and $H_y$ (TM-wave), whereas the wave corresponding to equation (40) has no components $H_y$ and $E_x$ (TE-wave). Relations (39) and (40) follow from (36) and (37) when only one component $j_x$ or $j_y$ is taken into account. However, a wave propagating at an angle to the $x$-axis is supported by two current components and has both tangential components of electric and magnetic fields. Naturally, the rotation of the coordinate system with the choice of the $x$ axis along or across the current translates equations (38) into (39) and (40), respectively. The results for dispersion of E-PP and H-PP along metallic films in dielectric, eq. (32) and (33), are presented in Fig.1, 2 and 3. Fig. 1 presents the dispersion for silver DEG film. In the plasmon resonance region, there are sharp transitions from very slow to very fast PP. The fine structure of spectral lines is shown in Fig. 2. For this purpose, losses were increased 10 and 100 times. Two deceleration peaks alternate through a narrow region of a fast wave with almost zero deceleration. The losses themselves for Fig. 2 are shown in Fig. 3. Since the branch $k'_x > 0$ is taken, that the region $k''_x < 0$ corresponds to backward PP.

For tensor surface conductivity we obtain generalized equations

$$j_{1x}\left[1+\frac{\sigma_{xx}(k_0^2\tilde{\varepsilon}-k_x^2)}{2\omega\varepsilon_0\tilde{\varepsilon}k_z}-\frac{\sigma_{xy}k_xk_y}{2\omega\varepsilon_0\tilde{\varepsilon}k_z}\right]+j_{1y}\left[\frac{\sigma_{xy}(k_0^2\tilde{\varepsilon}-k_y^2)}{2\omega\varepsilon_0\tilde{\varepsilon}k_z}-\frac{\sigma_{xx}k_xk_y}{2\omega\varepsilon_0\tilde{\varepsilon}k_z}\right]=0, \quad (41)$$

$$j_{1x}\left[\frac{\sigma_{xy}(k_0^2\tilde{\varepsilon}-k_x^2)}{2\omega\varepsilon_0\tilde{\varepsilon}k_z}-\frac{\sigma_{yy}k_xk_y}{2\omega\varepsilon_0\tilde{\varepsilon}k_z}\right]+j_{1y}\left[1+\frac{\sigma_{yy}(k_0^2\tilde{\varepsilon}-k_y^2)}{2\omega\varepsilon_0\tilde{\varepsilon}k_z}-\frac{\sigma_{xy}k_xk_y}{2\omega\varepsilon_0\tilde{\varepsilon}k_z}\right]=0. \quad (42)$$

The equality of the determinant of this homogeneous system of equations to zero $\hat{A}\mathbf{j}_1=0$ gives DE $A_{11}A_{22}-A_{12}A_{21}=0$. The elements in the DE are simplified with a diagonal conductivity tensor as $A_{11}=1+\sigma_{xx}(k_0^2\tilde{\varepsilon}-k_x^2)/(2\omega\varepsilon_0\tilde{\varepsilon}k_z)$, $A_{12}=-\sigma_{xx}k_xk_y/(2\omega\varepsilon_0\tilde{\varepsilon}k_z)$, $A_{22}=1+\sigma_{yy}(k_0^2\tilde{\varepsilon}-k_y^2)/(2\omega\varepsilon_0\tilde{\varepsilon}k_z)$. Further simplification occurs if the plasmon moves along one of the axes. Let, for example, $k_y=0$. Then the DE for the plasmon along the $x$ axis has the form

$$\left[1+\frac{\sigma_{xx}(k_0^2\tilde{\varepsilon}-k_x^2)}{2\omega\varepsilon_0\tilde{\varepsilon}k_z}\right]\left(1+\frac{\sigma_{yy}k_0^2}{2\omega\varepsilon_0 k_z}\right)=0. \quad (43)$$

It splits into two:

$$k_x^2=k_0^2\tilde{\varepsilon}\left(1-4\tilde{\varepsilon}/(\eta_0\sigma_{xx})^2\right), \quad (44)$$

$$k_x^2=k_0^2\tilde{\varepsilon}\left(1-(\eta_0\sigma_{yy}/2)^2/\tilde{\varepsilon}\right), \quad (45)$$

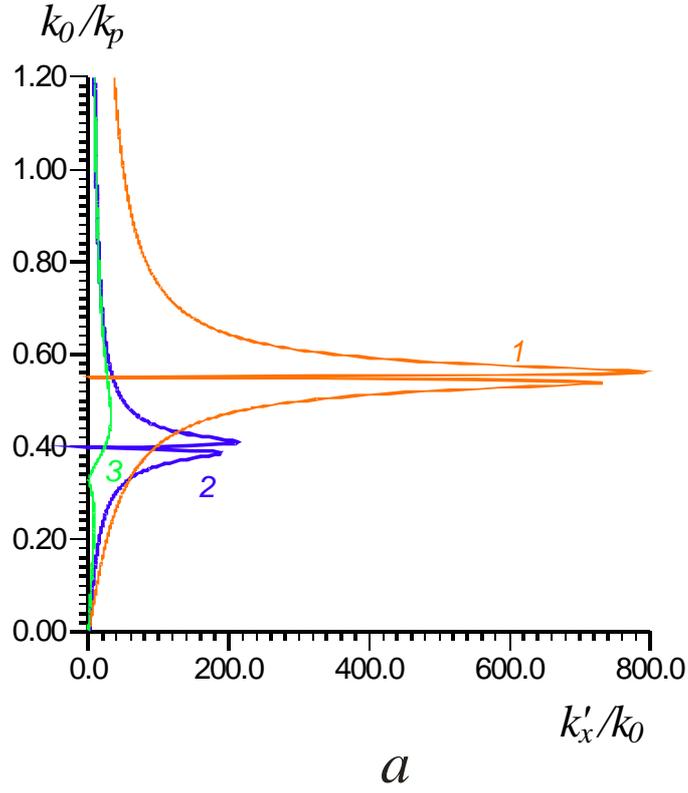

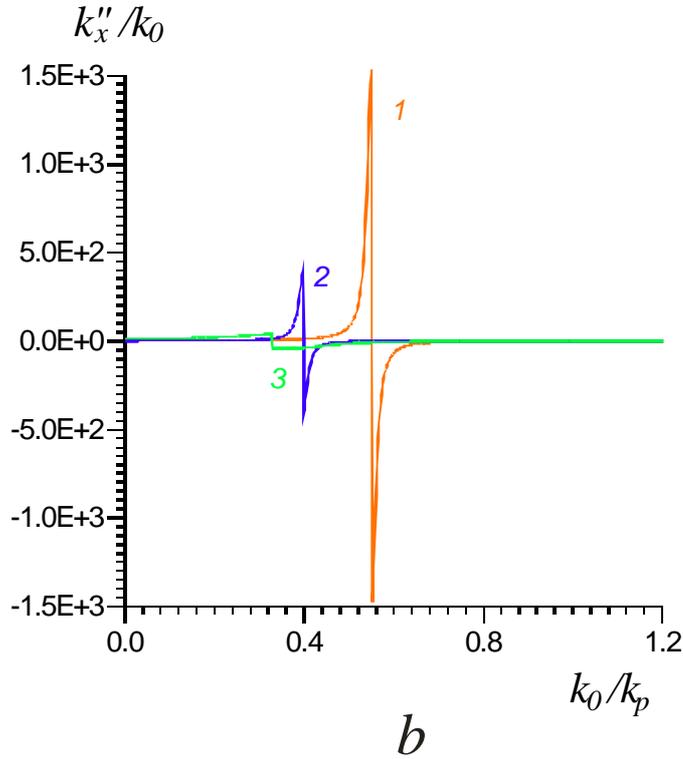

Figure 2. Normalized dispersion $k'_x/k_0$ (*a*) and normalized losses $k''_x/k_0$ (*b*) for E-PP along film $h$=2 nm with $\varepsilon_L = 9.3$, $\omega_p = 1.57 \cdot 10^{16}$ in dielectric $\tilde{\varepsilon} = 6 - 0.001i$ (curve 1) and $\tilde{\varepsilon} = 3 - 0.001i$ (2,3). Curves 1, 2 are plotted for $\omega_{co} = 3.56 \cdot 10^{14}$, curve 3 – for $\omega_{co} = 3.56 \cdot 10^{15}$ (frequencies in Hz)

each of which defines a wave in two opposite directions. Each of the equations (38) and (43) also corresponds to four waves. Note that the above equations are conveniently written using dimensionless (normalized) conductivity $\hat{\tilde{\xi}} = \eta_0 \hat{\sigma}$. In this case $\hat{\sigma}/(\omega \varepsilon_0) = \hat{\tilde{\xi}}/k_0$.

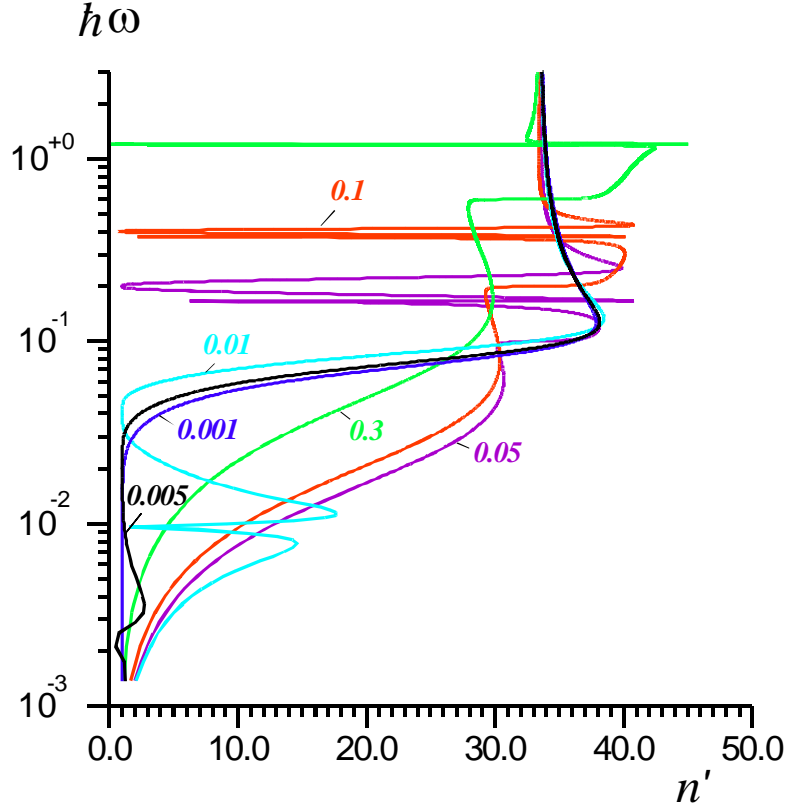

Figure 3. Dispersion of E-PP along graphene (dependence $\omega\hbar$, eV on deceleration $n' = k'_x/k_0$) for different values of the chemical potential (eV), $\omega_{co} = 10^{12}$ Hz

The results $n = n' - in'' = k_x/k_0$ for PP in graphene in vacuum, taking into account SD, are shown in Fig. 3–6. They demonstrate single and multiple transitions from fast PP to slow and vice versa, and for sections of fast PP, the dispersion curves may tend to zero $n'$ values of deceleration, and their drawing requires a very small frequency step. Each such transition is accompanied by a change in the sign of losses, which occurs at points $n = 1$. The existence of inverse PP can be confirmed by calculating the Poynting vector. It can be seen from the figures that the areas of reverse losses do not strictly correspond to the areas of anomalous negative dispersion. The use of the group velocity for the classification of forward and reverse

PP is possible only with low dissipation, i.e. in the regions with $n' \approx 1$, $n'' \ll 1$. At high decelerations, the dissipation is large, and the PP is forward. The transition to backward PP on curves with a negative slope occurs with a small dissipation $n \sim 1$.

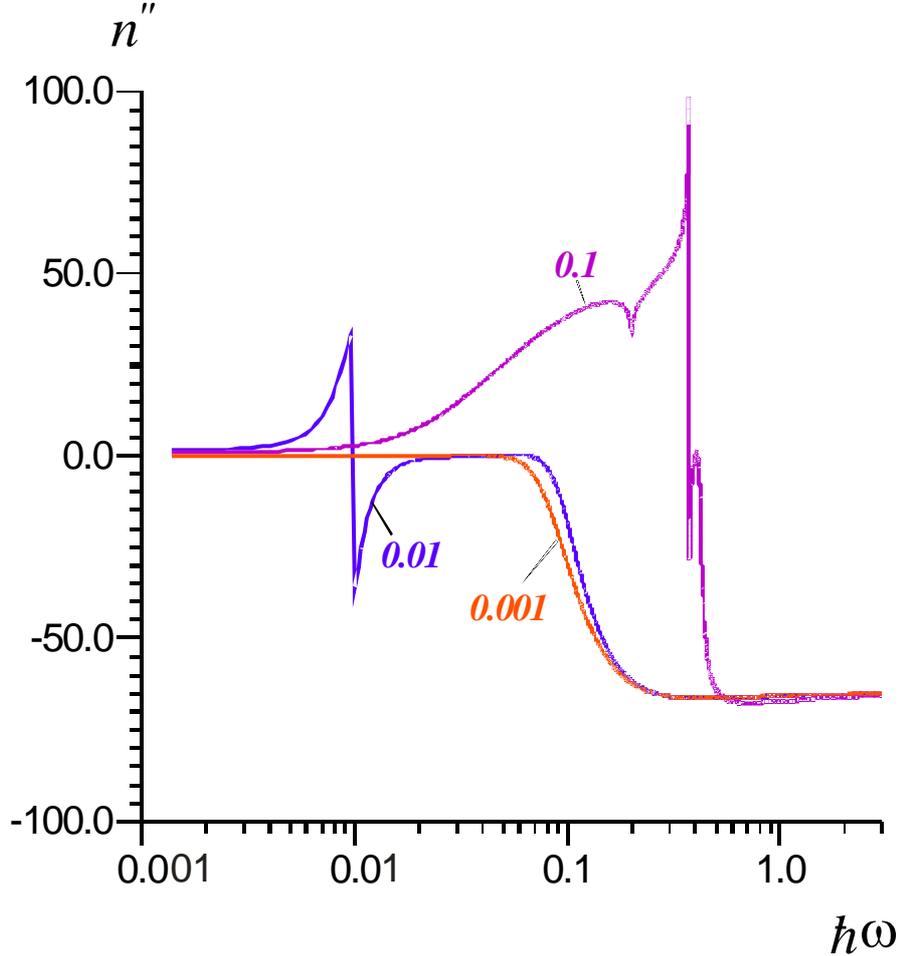

Figure 4. Dependence of losses $n'' = k_x''/k_0$ on $\omega \hbar$ (eV) for E-plasmon along graphene at different values of chemical potential (eV), $\omega_{co} = 10^{12}$ Hz

In this region, the PP goes at the speed of light, its energy is transferred in a vacuum, and the frequency, where there $n=1$ is a boundary between the gliding inflow and leakage PP [55].

The real part of $k_z$ for the incident current which exciting the structure in the GF approach must be positive. This is the radiation condition. Such current radiates waves in both directions $\pm z$. In dissipative films, the square root in equations (39) and (40) does not satisfy this condition. The fulfillment of the radiation condition is possible only in the case of an active film. The dissipation corresponds to $\omega_{co} > 0$.

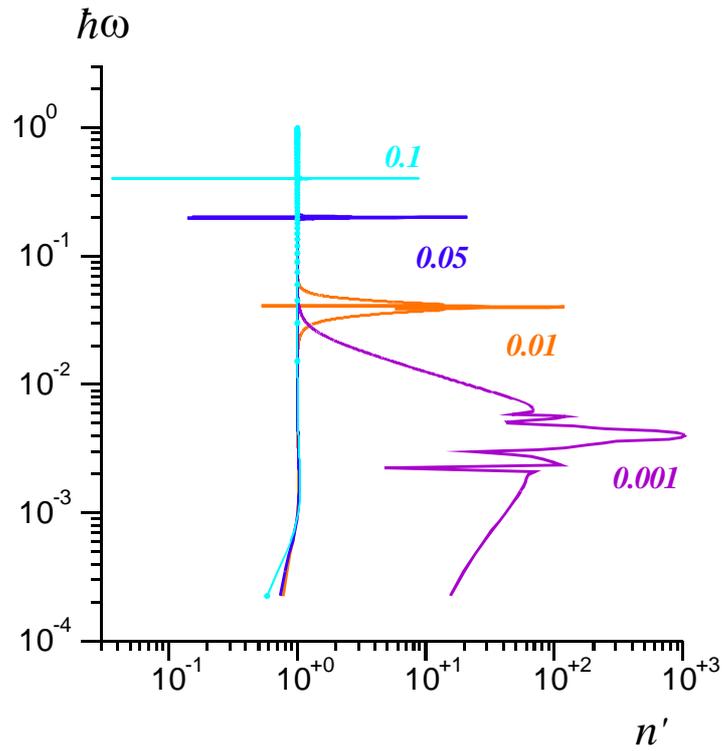

Рис. 5. Dispersion of H-PP along graphene (dependence $\omega\hbar$, eV on deceleration $n' = k'_x / k_0$) for different values of the chemical potential (eV), $\omega_{co} = 10^{12}$ Hz

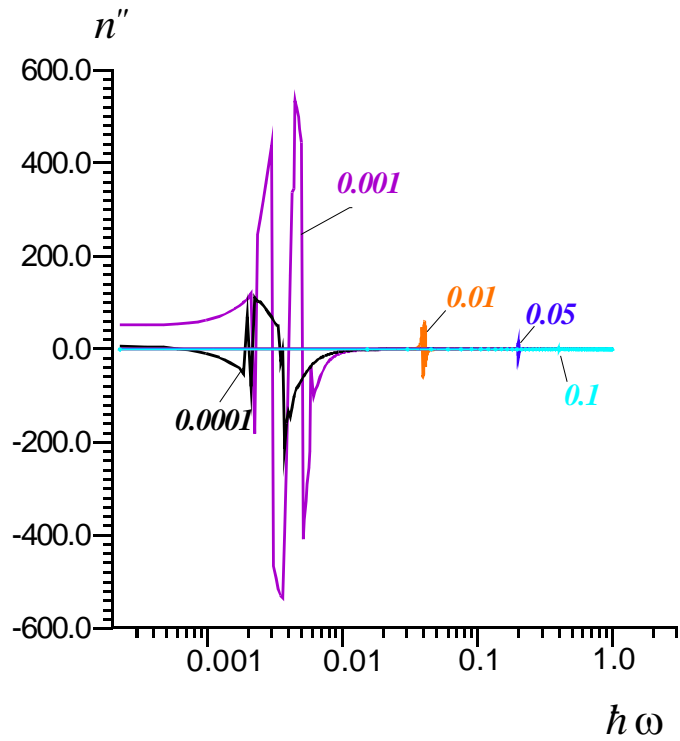

Figure 6. Losses $n'' = k''_x / k_0$ versus $\omega\hbar$ (eV) for H-PP along graphene at different values of chemical potential (eV), $\omega_{co} = 10^{12}$ Hz

For such PP the condition $k'_x k''_x > 0$ corresponds to the forward wave, and the condition $k'_x k''_x < 0$ – to backward wave. The slow PP corresponds to condition $\sigma'_{xx} \ll |\sigma''_{xx}|$, and for E-PP $\sigma''_{xx} < 0$ (inductive conductivity) means forward PP, and $\sigma''_{xx} > 0$ (capacitive conductivity) means backward PP. If the meta-surface has the structure of periodic two-dimensional conductive elements on a multilayer substrate, then appropriate GFs of layered structures should be used. They are easily obtained using transmission matrices [55]. In general, these are fourth-order matrices, known in optics as the Berreman matrix [56]. When stitching the Fourier components of 4-vectors $u = (e_x, \eta_0 h_y, -e_y, \eta_0 h_x)$, the normalized Berreman matrix for a graphene sheet on the dielectric layer has the form

$$\hat{T}_\sigma = \begin{bmatrix} 1 & 0 & 0 & 0 \\ \xi_{xy} & 1 & -\xi_{yy} & 0 \\ 0 & 0 & 1 & 0 \\ \xi_{xx} & 0 & -\xi_{xy} & 1 \end{bmatrix}, \quad (41)$$

and the matrix for the substrate is formed as

$$\hat{T}(d) = \begin{bmatrix} \hat{a}^e(d) + \hat{a}^h(d) & \hat{0} \\ \hat{0} & \hat{a}^e(d) + \hat{a}^h(d) \end{bmatrix}. \quad (42)$$

It contains zero two-dimensional matrices $\hat{0}$ and two-dimensional matrices

$$\hat{a}^{e,h}(d) = \begin{bmatrix} \cos(k_z d) & i\rho_{e,h} \sin(k_z d) \\ iy_{e,h} \sin(k_z d) & \cos(k_z d) \end{bmatrix}.$$

Here $k_z = \sqrt{k_0^2 \varepsilon - k_x^2 - k_y^2}$, $\rho_e = k_z/(k_0 \varepsilon)$ and $\rho_h = k_0/k_z$ are normalized wave impedances of E-wave and H-wave. PP is often considered on the surface of metamaterials [48–50]. In periodic layered structure with DPs $\varepsilon_1$, $\varepsilon_2$, thicknesses $t_1$, $t_2$ and the period $t = t_1 + t_2$ the low frequency Rytov homogenization reads [57] $\varepsilon_{xx} = \varepsilon_{yy} = (\varepsilon_1 t_1 + \varepsilon_2 t_2)/t$ and $\varepsilon_{zz} = t(t_1/\varepsilon_1 + t_2/\varepsilon_2)^{-1}$. The homogenization with taking into account SD gives [49,50]:

$$\varepsilon_{xx} = \varepsilon_{yy} = k_0^{-2}\left[k^2/2 + k_{1z}^2 t^{-2}(t_1^2 + t_1 t_2) + k_{2z}^2 t^{-2}(t_2^2 + t_1 t_2)\right], \quad (43)$$

$$\varepsilon_{zz} = k_0^{-2} k^2 \left\{1 - \frac{k_{1z}^2 t^{-2}(t_1^2 + t_1 t_2 \varepsilon_2/\varepsilon_1) + k_{2z}^2 t^{-2}(t_2^2 + t_1 t_2 \varepsilon_1/\varepsilon_2)}{k^2/2 + k_{1z}^2 t^{-2}(t_1^2 + t_1 t_2) + k_{2z}^2 t^{-2}(t_2^2 + t_1 t_2)}\right\}^{-1}. \quad (44)$$

Here $k_{1z}^2 = k_0^2\varepsilon_1 - k^2$, $k_{2z}^2 = k_0^2\varepsilon_2 - k^2$, $k^2 = k_x^2 + k_y^2$. It is possible the Diakonov PPs propagation at the interface of such metamaterial with vacuum [48].

Recently, there has been an increased interest in magnetoplasmons in thin conductive films. Such a film is described by tensor DP. The problem can be solved strictly [58]. Here we consider an approximation of 2D conductivity when the magnetic field is directed along the propagation of the magnetoplasmon ($x$-axis). We have $\varepsilon_{xx}(\omega) = \varepsilon = \varepsilon_L - \omega_p^2/(\omega^2 - i\omega\omega_c)$, $\varepsilon_{yy}(\omega) = \varepsilon_{zz}(\omega) = \varepsilon_L - \omega_p^2/(\omega^2 - \omega_H^2 - i\omega\omega_c)$, $\varepsilon_{yz}(\omega) = -\varepsilon_{zy}(\omega) = -ib = -i\omega_p^2\omega_H\omega^{-1}/(\omega^2 - \omega_H^2 - i\omega\omega_{co})$, $\omega_H = \mu_0 H_{0z} e/m_e$ is cyclotron frequency, $\sigma_{xx} = ik_0 t\eta_0^{-1}(\varepsilon - 1)$, $\sigma_{xy} = \sigma_{yx} = \sigma_{xz} = \sigma_{zx} = 0$, $\sigma_{yz} = -\sigma_{zy} = k_0 t\eta_0^{-1}b$, $\sigma_{yy} = \sigma_{zz} = ik_0 t\eta_0^{-1}(\varepsilon_{yy} - 1)$. Let enter normalised conductivity $\hat{\xi} = \hat{\sigma}/\eta_0$. For a wave along the $x$ axis, $k_y=0$, and we should take $g_{xy} = g_{yx} = g_{yz} = g_{zy} = 0$, $g_{xz} = k_x/(2k_0)$, $g_{xx} = -\tilde{k}_z/(2k_0)$, $g_{yy} = -k_0/(2\tilde{k}_z)$, $g_{zz} = -k_x^2/(2k_0\tilde{k}_z)$, and the equations for spectral amplitudes

$$E_x(1 - g_{xx}\xi_{xx}) = g_{xz}\xi_{zy}E_y + g_{xz}\xi_{zz}E_z,$$
$$E_y(1 - g_{yy}\xi_{yy}) = g_{yy}\xi_{yz}E_z, \tag{45}$$
$$E_z(1 - g_{zz}\xi_{zz}) = g_{zz}\xi_{zy}E_y + g_{zx}\xi_{xx}E_x,$$

from which we have the DE

$$(1 - g_{xx}\xi_{xx}) = \psi(\omega, k_x) = g_{zx}\xi_{xx}\frac{g_{yy}g_{xz}\xi_{yz}\xi_{zy} + g_{xz}\xi_{zz}(1 - g_{yy}\xi_{yy})}{(1 - g_{yy}\xi_{yy})(1 - g_{zz}\xi_{zz}) - g_{yy}g_{zz}\xi_{yz}\xi_{zy}}, \tag{46}$$

or $n = k_x/k_0 = \sqrt{1 - (2(\psi - 1)/\xi_{xx})^2}$. The 2D approximation means neglecting the transverse current, but here we have taken it into account. Otherwise, DE does not describe the variance. You should define $\sigma_{zz}$ and $\sigma_{zy}$. For a two–dimensional DEG film the continuity relation $\nabla_\tau \cdot \mathbf{J}_\tau = -i\omega\varsigma_0$ holds for surface charge density $\varsigma_0$. For the volumetric current density the continuity relation $\nabla \cdot \mathbf{j} = \nabla_\tau \cdot \mathbf{J}_\tau / t + \partial_z j_z = 0$ also holds (there are no volumetric charges). The dependence of the transverse component on $z$ should have the form $j_z(z) = j_{0z} + j_{1z} z/t$, since the ratio must be fulfilled, and the component $E_z(z)$ due to the surface nature of the current should have a jump when passing through the film, i.e. have a dependence $E_z(z) = E_{z0} 2z/t$. For the specified jump we have $E_z(t/2) - E_z(-t/2) = \varsigma_0 / \varepsilon_0 = 2 E_{z0}/t$, moreover $j_{1z} = i\omega \varsigma_0$, $j_{0z} = i\omega \varepsilon_0 \varepsilon_{zy} E_y$. Writing down $j_z(z)/t = \sigma_{zz} E_z(z) + \sigma_{zy} E_y(z)$, we get $\sigma_{zz} = i\omega \varepsilon_0 (\varepsilon_{zz} - 1) t$, $\sigma_{zy} = i\omega \varepsilon_0 \varepsilon_{zy} t$. The results of calculating the dispersion of such a magnetoplasmons are shown in Fig. 7.

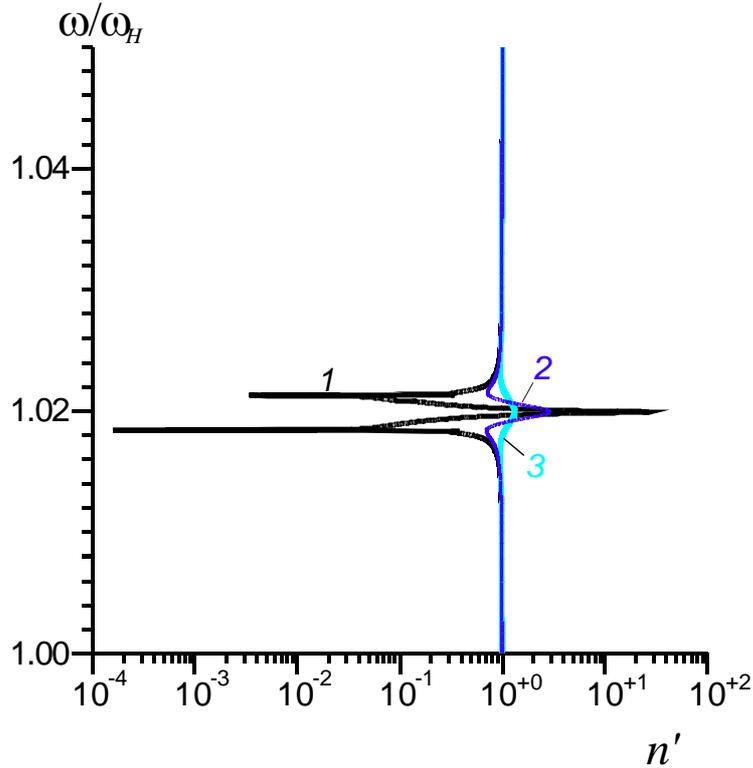

Рис. 7. Дисперсия магнитоплазмона в 2DEG $n$-InSb $t=100$ нм вдоль оси $x$ в зависимости от замедления $n' = \mathrm{Re}(k_x/k_0)$ при различных частотах столкновений $\omega_c = 1.6$ ГГц (кривая 1), $\omega_c = 16$ ГГц (2) и $\omega_c = 50$ ГГц. $\omega_p = 56$ ТГц, $\omega_H = 14.6$ ТГц

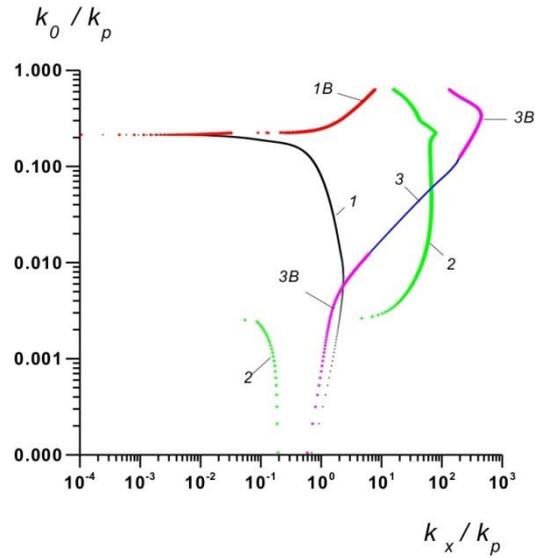

Figure 8. Dispersion of the H-polariton on the metasurface in the form of a dielectric layer of $t = 50$ nm thickness with a DP $\tilde{\varepsilon} = 3 - 0.01$ on a metal half-space (curve 1) and the same layer with periodic doping. Curve 2 – four periods for the thickness of the layer with silver balls $r = 8.33$ nm; curve 3 - four periods for the thickness of the layer with silver balls $r = 6.25$ nm. The symbol $B$ to the right of the number means the backward wave

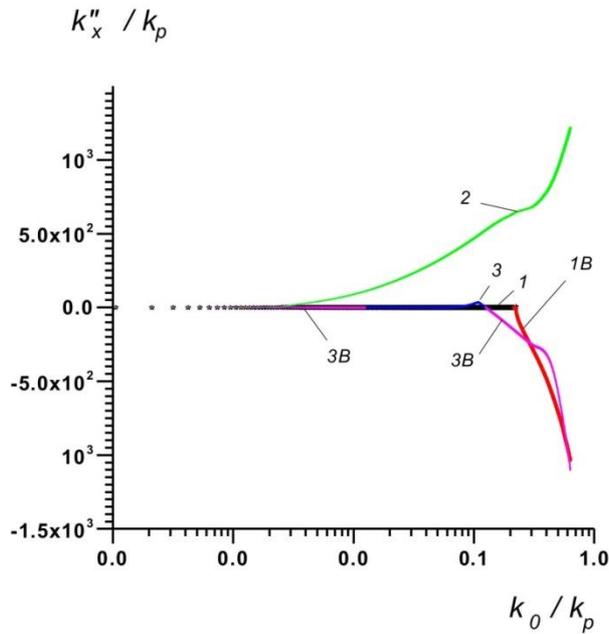

Figure 9. Normalized losses of H-PP on the metasurfaces corresponding to Fig. 8. The symbol $B$ means the backward PP

Along a weakly dissipative metallic surface the maximum deceleration and loss of PP approximately correspond to the point $\omega'_s \approx \omega_s$, $\varepsilon' \approx -1$, while $\varepsilon'' \ll |\varepsilon'|$ and $k'_x(\omega_s) = k_0(1+\varepsilon''/2)/\sqrt{2\varepsilon''}$, $k''_x(\omega_s) = k_0(1-\varepsilon''/2)/\sqrt{2\varepsilon''}$, i.e. $k'_x \approx k''_x \approx k_0/\sqrt{2\varepsilon''}$. For metal $\varepsilon'' = \omega_p^2 \omega_c /(\omega_s^3 + \omega_s \omega_c^2) \approx (\varepsilon_L+1)^{3/2} \omega_c/\omega_p$, and when $\omega_c/\omega_p = 10^{-3}$ we get $\varepsilon'' \sim 0.03$. The maximum losses occur at the frequency of the plasmon resonance (see Fig. 1,2), but a small detuning down allows you to obtain sufficiently slow plasmons with acceptable losses. In the case of a multilayer medium with an impedance surface, we have implicit equations. Let the normalized input impedance such matasurface at $z=0$ has the form $\rho_{in} = \rho' + i\rho''$. Then $k_x/k_0 = \sqrt{1-\rho_{in}^2(k_x,k_0)}$ for E-PP. If $\rho''^2(k_x) \gg \rho'^2(k_x) > 1$, the E-PP is slow. In this case $k_x/k_0 \approx \sqrt{1+\rho''^2}(1-i\rho'\rho''/(1+\rho''^2))$, and E-PP is backward if $\rho'' < 0$, i.e. for capacitive input impedance. For H-PP we have $k_x/k_0 = \sqrt{1-\rho_{in}^{-2}(k_x,k_0)}$, and H-PP is backward if $\rho'' > 0$, i.e. for inductive input impedance.

Let's consider a metasurface in the form of a thin dielectric layer with DP, $\tilde{\varepsilon}$ into which metatoms in the form of metal balls with nanoscale radius $r$ are embedded. We use an approximation of effective medium in the form of Bruggeman or J.K.M. Garnett (Wiener-Wagner) type. Periodic and non-periodic implementations may be considered. Figures 8,9 show the results for the initial introduction of metal balls into the dielectric with $\tilde{\varepsilon} = 3 - 0.001i$ (SiO2). We used two cases with smallest distance between balls 2r and r. In the first case, the metal filling factor $c_2 = 0.155$ is close to the leakage threshold, whereas in the second case it is small: $c_2 = 0.0655$. We have used the Bruggeman effective field formula and have considered the dielectric $SiO_2$ film with thickness $t = 50$ nm and $t = 100$ nm, in which there are four periods of layers with a two-dimensional periodic arrangement of metal balls. In the first case, $r = 8.3$ nm, and in the second case, $r = 6.2$ nm. The metal is described by the parameters $\varepsilon_L = 9 - 0.01i$, $\omega_p = 10^{16}$, $\omega_c = 10^{14}$ Hz. Figure 8 shows the results for the dispersion of H-PP on the metasurface in the form of a $SiO_2$ film with a thickness of 50 nm located on the metal, as well as a film with a thickness of 100 nm doped with metal balls according to the first and second variants. Fig. 9 shows the normalized losses of H-polaritons. The solution of the DE is obtained by the iteration method. The branches in Fig. 8,9 for backward are marked with the symbol $B$.

## 4. Conclusions

In this paper we have considered the planar metasurfaces with thin conducting films, the electrodynamic tensor GFs, the dispersion equations of surface PP. We dealt mainly with an electrodynamically thin metasurfaces. Some methods of introduction and calculation of effective surface conductivities are given, as well as a number of results for dispersion and losses of plasmons. The conditions for the existence of slow, fast and backward plasmon waves are determined. Changing the type of input impedance from inductive to capacitive leads to a transition from forward to backward wave. The doping of the metasurface layer with nanoparticles or the execution of a certain structure on it allows you to change the character of the polariton from forward to backward, which is one of the ways to create focusing structures on metasurfaces. For thin metal films, such structures can be strips with holes and inclusions from air bubbles.

**Acknowledgements.** This research was supported by the Ministry of Education and Science of Russia within the framework of the state task (project No. FSRR-2023-0008) and by the grant from the Russian Science Foundation No. 21-19-00226.

**Disclosure statement**

No potential conflict of interest was reported by the author.

**ORCID**

*Davidovich@http://orcid.org/0000-0001-8706-8523*